\documentclass[sigconf]{acmart}

\usepackage{booktabs} 
\usepackage{subfigure}
\usepackage[linesnumbered,ruled,vlined]{algorithm2e}
\usepackage{algpseudocode}
\usepackage{comment}

\setcopyright{none}

\copyrightyear{2017} 
\acmYear{2017} 
\setcopyright{usgovmixed}
\acmConference{SC17}{November 12--17, 2017}{Denver, CO, USA}\acmPrice{15.00}\acmDOI{10.1145/3126908.3126912}
\acmISBN{978-1-4503-5114-0/17/11}

\begin{document}
\title{Scaling Deep Learning on GPU and Knights Landing clusters}

\author{Yang You}
\affiliation{%
  \institution{Computer Science Division}
  \streetaddress{UC Berkeley}
}
\email{youyang@cs.berkeley.edu}

\author{Ayd\i n Bulu\c{c}}
\affiliation{%
  \institution{Computer Science Division}
  \streetaddress{LBNL and UC Berkeley}
}
\email{abuluc@lbl.gov}

\author{James Demmel}
\affiliation{%
  \institution{Computer Science Division}
  \streetaddress{UC Berkeley}
  }
\email{demmel@cs.berkeley.edu}


\begin{abstract}
The speed of deep neural networks training has become a big bottleneck of deep learning research and development. For example, training GoogleNet by ImageNet dataset on one Nvidia K20 GPU needs 21 days \cite{iandola2016firecaffe}.
To speed up the training process, the current deep learning systems heavily rely on the hardware accelerators. However, these accelerators have limited on-chip memory compared with CPUs. To handle large datasets, they need to fetch data from either CPU memory or remote processors.  
We use both {\bf self-hosted} Intel Knights Landing (KNL) clusters and multi-GPU clusters as our target platforms. 
From an algorithm aspect, current distributed machine learning systems \cite{dean2012large} \cite{li2014scaling} are mainly designed for cloud systems. These methods are asynchronous because of the slow network and high fault-tolerance requirement on cloud systems. 
We focus on Elastic Averaging SGD (EASGD) \cite{zhang2015deep} to design algorithms for HPC clusters.
Original EASGD \cite{zhang2015deep} used round-robin method for communication and updating. The communication is ordered by the machine rank ID, which is inefficient on HPC clusters. 

First, we redesign four efficient algorithms for HPC systems to improve EASGD's poor scaling on clusters. Async EASGD, Async MEASGD, and Hogwild EASGD are faster \textcolor{black}{than} their existing counterparts (Async SGD, Async MSGD, and Hogwild SGD, resp.) in all the comparisons. Finally, we design Sync EASGD, which ties for the best performance among all the methods while being deterministic.
In addition to the algorithmic improvements, we use some system-algorithm codesign techniques to scale up the algorithms.
By reducing the percentage of communication from 87\% to 14\%, our Sync EASGD achieves 5.3$\times$ speedup over original EASGD on the same platform. We get 91.5\% weak scaling efficiency on 4253 KNL cores, which is higher than the state-of-the-art implementation. 
\end{abstract}

%
%



\maketitle

\section{Introduction}
For deep learning applications, larger datasets and bigger models lead to significant improvements in accuracy \cite{amodei2015deep}. However, the computational power for training deep neural networks has become a big bottleneck. The current deep networks require days or weeks to train, which makes real-time interaction impossible. For example, training ImageNet by GoogleNet on one Nvidia K20 GPU needs 21 days \cite{iandola2016firecaffe}. Moreover, the neural networks are rapidly becoming more and more complicated. For instance, state-of-the-art Residual Nets have 152 layers \cite{he2016deep} while the best networks four years ago (AlexNet \cite{krizhevsky2012imagenet}) had only 8 layers. 
To speed up the training process, the current deep learning systems heavily rely on hardware accelerators because they can provide highly fine-grained data-parallelism (e.g. GPGPUs) or fully-pipelined instruction-parallelism (e.g. FPGA). However, these accelerators have limited on-chip memory compared with CPUs. To handle big models and large datasets, they need to fetch data from either CPU memory or remote processors at runtime. Thus, reducing communication and improving scalability are critical issues for distributed deep learning systems. 

To explore architectural impact, in addition to multi-GPU platform, we choose the Intel Knights Landing (KNL) cluster as our target platform. KNL is a self-hosted chip with more cores than CPUs (e.g. 68 or 72 vs 32). Compared with its predecessor Knights Corner (KNC), KNL significantly improved both computational power (6 Tflops vs 2 Tflops for single precision) and memory bandwidth efficiency (450 GB/s vs 159 GB/s for STREAM benchmark). Moreover, KNL introduced MCDRAM and configurable NUMA, which are highly important for applications with complicated memory access patterns. 
We design communication-efficient deep learning methods on GPU and KNL clusters for better scalability.

Algorithmically, current distributed machine learning systems \cite{dean2012large} \cite{li2014scaling} are mainly designed for cloud systems. These methods are asynchronous because of the slow network and high fault-tolerance requirement on cloud systems. A typical HPC cluster's bisection bandwidth is 66.4 Gbps (NERSC Cori) while the data center's bisection bandwidth is around 10 Gbps (Amazon EC2). However, as mentioned before, the critical issues for current deep learning system are speed and scalability. Therefore, we need to select the right method as the starting point.
Regarding algorithms, we focus on Elastic Averaging SGD (EASGD) method since it has a good convergence property \cite{zhang2015deep}.
\textcolor{black}{Original EASGD used a round-robin method for communication. The communication is ordered by the machine rank ID. At any moment, the master can interact with just a single worker. The parallelism is limited to the pipeline among different workers. Original EASGD is inefficient on HPC systems.} 

First, we redesign four efficient distributed algorithms to improve EASGD's poor scaling on clusters. By changing the round-robin style to parameter-server style, we got Async EASGD. After adding momentum \cite{sutskever2013importance} to Async EASGD, we got Async MEASGD. Then we combine Hogwild method and EASGD updating rule to get Hogwild EASGD. Async EASGD, Async MEASGD, and Hogwild EASGD are faster \textcolor{black}{than} their existing counterparts (i.e. Async SGD, Async MSGD, and Hogwild SGD, resp.). Finally, we design Sync EASGD, which ties for the best performance among all the methods while being deterministic (Figure \ref{fig:compare}).
Besides the algorithmic refinements, the system-algorithm codesign techniques are important for scaling up deep neural networks. The techniques we introduce include: (1) using single-layer layout and communication to optimize the network latency and memory access, (2) using multiple copies of weights to speedup the gradient descent, and (3) partitioning the KNL chip based on data/weight size and reducing communication on multi-GPU systems. 
By reducing the communication percent from 87\% to 14\%, our Sync EASGD achieves 5.3$\times$ speedup over original EASGD on the same platform. Using ImageNet dataset to train GoogleNet on 2176 KNL cores, the weak scaling efficiency of Intel Caffe is 87\% while our implementation is 92\%. Using ImageNet to train VGG on 2176 KNL cores, the weak scaling efficiency of Intel Caffe is 62\% while our implementation is 78.5\%.
\textcolor{black}{To highlight the difference between existing methods and our methods, we list our three major contributions: }
 
\textcolor{black}{(1) {\bf Sync EASGD and Hogwild EASGD algorithms}. We have documented our process in arriving at these two algorithms, which ultimately perform better than existing methods. 
The existing EASGD uses round-robin updating rule. We refer to the existing method as Original EASGD. We first changed the round-robin rule to parameter-server rule to arrive at Async EASGD. The difference between Original-EASGD and Async-EASGD is that the updating rule of Original-EASGD is ordered while Async-EASGD is unordered. Adding momentum to that we arrived at Async MEASGD. Neither Async EASGD nor Async MEASGD were significantly faster than Original EASGD (Figure \ref{fig:compare}).}

\textcolor{black}{In both Original-EASGD and Async-EASGD, the master only communicates with one worker at a time. Then we relaxed this requirement to allow the master to communicate with multiple workers at a time to get Hogwild EASGD. The master first receives multiple weights from different workers. The master then processes these weights by the Hogwild (lock-free) Updating rule. We observe that the lock-free Hogwild makes Hogwild EASGD run much faster than Original EASGD. For the convex case, we can prove the algorithm is safe and faster under some assumptions\footnote{\href{https://www.cs.berkeley.edu/~youyang/HogwildEasgdProof.pdf}{https://www.cs.berkeley.edu/$\sim$youyang/HogwildEasgdProof.pdf}}.}

\textcolor{black}{We used tree reduction algorithm to replace round-robin rule to get Sync EASGD. Sync EASGD is much faster ($\Theta$(logP) vs $\Theta$(P)). This is highly important because deep learning researchers often need to tune many hyperparameters, which is exetremly time-consuming.
While not being one of our major contributions, we also documented the relative order of performance between intermediate algorithms we have considered. For instance, we observe that Async EASGD is faster than Async SGD and Async MEASGD is faster than Async MSGD.}

\textcolor{black}{(2) {\bf Algorithm-System Co-design for multi-GPU system}. After the algorithm-level optimization, we need to produce an efficient design on the multi-GPU system. We reduce the communication overhead by changing the data's physical locations. We also design some strategies to overlap the communication with the computation. After the algorithm-system co-design, our implementation achieves a 5.3x speedup over the Original EASGD.}

\textcolor{black}{(3) {\bf Use KNL cluster to speedup DNN training}. GPUs are good tools to train deep neural networks. However, we also want to explore more hardware options for deep learning applications. We choose KNL because it has powerful computation and memory units. Section \ref{sec:knl_optimization} describes the optimization for small dataset DNN training on KNL platform. In our experiments, using an 8-core CPU to train CIFAR-10 dataset takes 8.2 hours. However, CIFAR-10 is only 170 MB, which can not make full use of KNL's 384 GB memory. This optimization helps us to finish the training in 10 minutes. The optimization in Section \ref{sec:one_layer} is designed on KNL cluster, but it can be used on regular clusters.}

\section{Background}
We \textcolor{black}{describe} the Intel Knights Landing architecture, which is used in this paper. We \textcolor{black}{review} necessary background on deep learning for readers to understand this paper.

\subsection{Intel Knights Landing Architecture\label{sec:knl}}
Intel Knights Landing (KNL) Architecture is the latest version of Intel Xeon Phi. Compared with the previous version, i.e. Knights Corner (KNC), KNL has slightly more cores (e.g. 72 or 68 vs 60). Like KNC, each KNL core has 4 hardware threads and supports 512-bit instruction for SIMD data parallelism. 
The major distinct features of KNL include the following:

\begin{figure}[!t]
\centering
\includegraphics[width=3.2in]{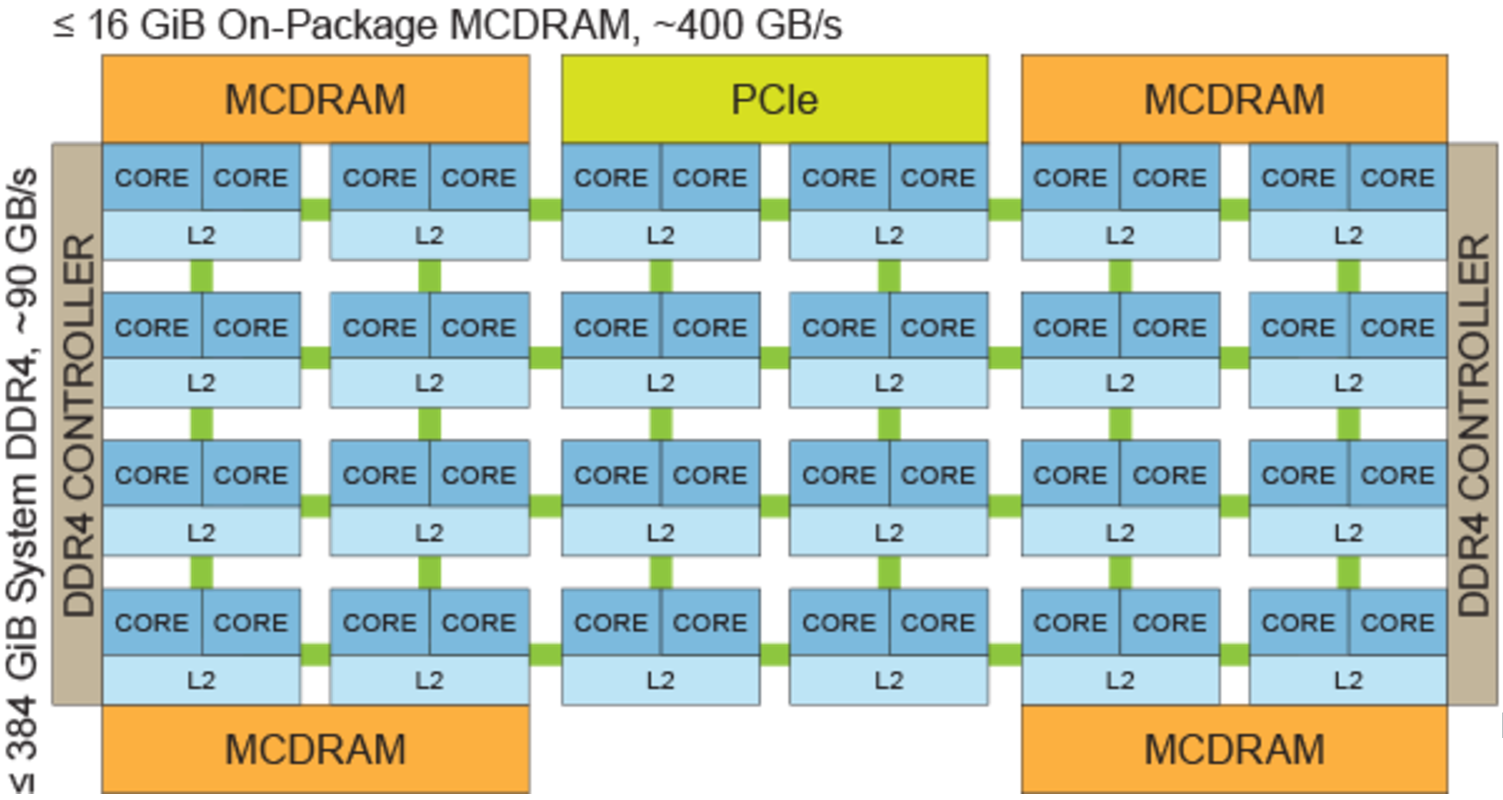}
\caption{The KNL Architecture. Our version has 68 cores.}
\label{fig:cnn}
\end{figure}

{\bf (1) Self-hosted Platform} The traditional accelerators (e.g. FPGA, GPUs, and KNC) rely on CPU for control and I/O management. For some applications, the transfer path like PCIE may become a bottleneck at runtime because the memory on accelerator is limited (e.g. 12 GB GDDR5 on one Nvidia K80 GPU). The KNL does not need a host. It is self-hosted by an operating system like CentOS 7.

\begin{figure*}[!t]
\centering
\includegraphics[width=5.6in]{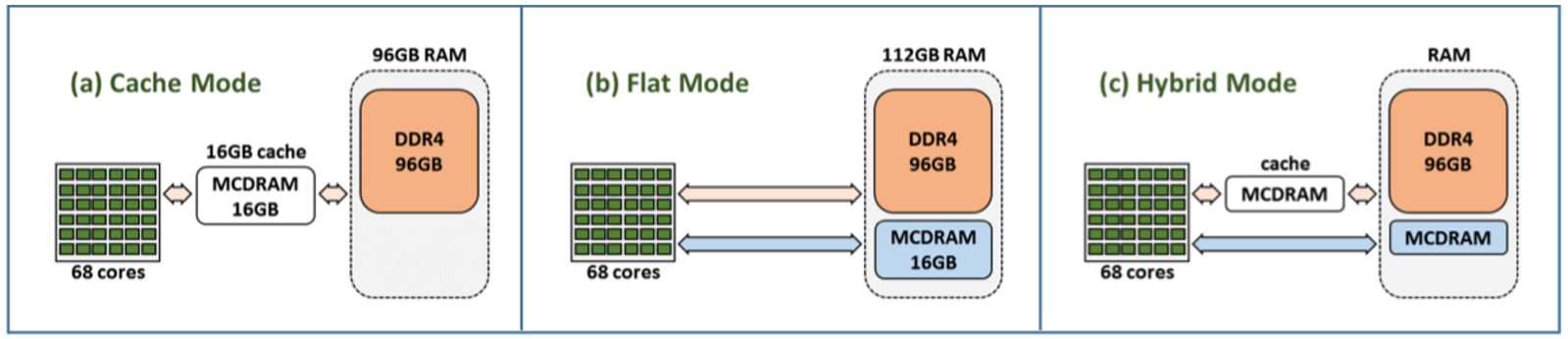}
\caption{The three modes of MCDRAM. Under Cache Mode, MCDRAM acts as the last-level cache. Under Flat Mode, MCDRAM is part of RAM. Under Hybird Mode, part of MCDRAM acts as the last-level cache, and the rest of it acts as RAM.}
\label{fig:mcdram}
\end{figure*}

{\bf (2) Better Memory} 
KNL's DDR4 memory size is much larger than that of KNC (384 GB vs 16 GB). Moreover, KNL is equipped with 16 GB Multi-Channel DRAM (MCDRAM). MCDRAM's measured bandwidth is 475 GB/s (STREAM benchmark). \textcolor{black}{The bandwidth of KNL's regular DDR4 is 90 GB/s.} MCDRAM has three modes: a) Cache Mode: KNL uses it as the last level cache; b) Flat Mode: KNL treats it as the regular DDR; c) Hybrid Mode: part of it is used as cache, the other is used as the regular DDR memory (Figure \ref{fig:mcdram}). 

{\bf (3) Configurable NUMA}
KNL supports all-to-all (A2A), quadrant/hemisphere (Quad/Hemi) and sub-NUMA (SNC-4/2) clustering modes of cache operation. For A2A, memory addresses are uniformly distributed across all tag directories (TDs) on the chip. For Quad/Hemi, the tiles are divided into four parts called quadrants, which are spatially local to four groups of memory controllers. Memory addresses served by a memory controller in a quadrant are guaranteed to be mapped only to TDs contained in that quadrant. Hemisphere mode functions the same way, except that the die is divided into two hemispheres instead of four quadrants. The SNC-4/2 mode partitions the chip into four quadrants or two hemispheres, and, in addition, expose these quadrants (hemispheres) as NUMA nodes. In this mode, NUMA-aware software can pin software threads to the same quadrant (hemisphere) that contains the TD and access NUMA-local memory. 

\subsection{DNN and SGD}
We focus on Convolutional Neural Networks (CNN) \cite{lecun1998gradient} in this section. Figure \ref{fig:cnn} is an illustration of CNN.
CNN is composed of a sequence of tensors. We refer to these tensors as weights. At runtime, the input of CNN is a picture $X$ (e.g. $X$ is stored as a 32-by-32 matrix in Figure \ref{fig:cnn}). After a sequence of tensor-matrix operations, the output of CNN is an integer $y$ (e.g. $y$ $\in$ $\{0, 1, 2, ..., 9\}$ in Figure \ref{fig:cnn}). The tensor-matrix operations can be implemented by dense matrix-matrix multiplication, FFT, dense matrix-vector multiplication, dense vector-vector add and non-linear transform (e.g. Tanh, Sigmoid, and ReLU \cite{goodfellow2016deep}).

Figure \ref{fig:cnn} is an example of hand-written image recognition. The input picture $X$ should be recognized as 3 by a human. If $y$ is 3, then the input picture is correctly classified by the CNN framework. To get the correct classification, we need to get a set of working weights. The weights needs to be trained by using the real-world datasets.
For simplicity, let us refer to the weights as $W$, and the training dataset as $\{X_i, y_i\}$, $i$ $\in$ $\{1, 2, ..., n\}$. $n$ is the number of training pictures. $y_i$ is the correct label for $X_i$. The training process includes three parts: 1) Forward Propagation, 2) Backward Propagation, and 3) Weight Update.

{\bf Forward Propagation} $X_i$ is passed from the first layer to the last layer of the neural network (left to right in Figure \ref{fig:cnn}). The output is the prediction of $X_i$'s label, which is referred to as $\tilde{y_i}$.

{\bf Backward Propagation} We get a numerical prediction error $E$ as the difference between ${y_i}$ and $\tilde{y_i}$. Then we pass $E$ from the last layer to the first layer to get the gradient of $W$, which is $\Delta W$.

{\bf Weight Update} We refine the CNN framework by updating the weight: $W$ $\leftarrow$ $W -  \eta \Delta W$, where $\eta$ is a number called the learning rate (e.g. 0.01).

We conduct the above three steps iteratively \textcolor{black}{over all the samples} until the model is optimized \textcolor{black}{(i.e. randomly picks a batch of samples at each iteration)}. This method is called Stochastic Gradient Descent (SGD) \cite{goodfellow2016deep}. Stochastic means we randomly pick a batch of $b$ pictures at each iteration. Usually $b$ is an integer chosen from 16 to 2048. If $b$ is too large, SGD's convergence rate usually will decrease \cite{li2014efficient}. 

\begin{figure}[!t]
\centering
\includegraphics[width=3.2in]{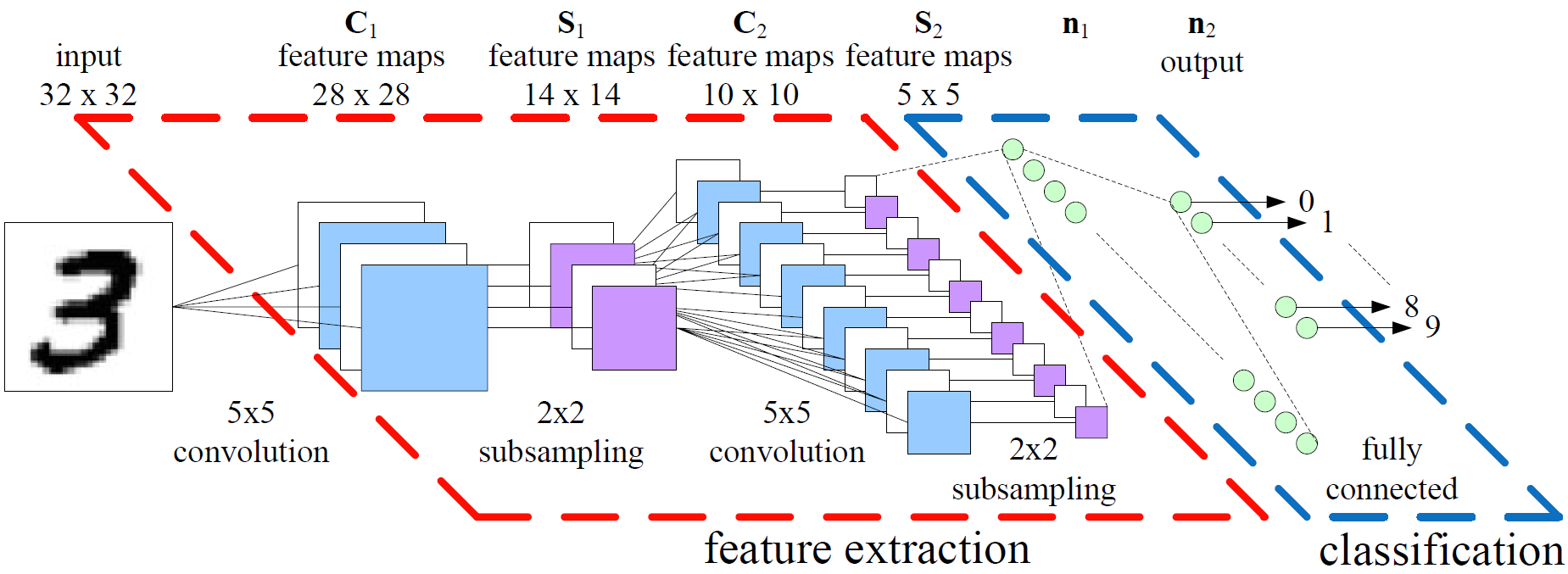}
\caption{This figure \cite{peemen2011efficiency} is an illustration of Convolutional Neural Network.}
\label{fig:cnn}
\end{figure}

\subsection{Data Parallelism and Model Parallelism}
\begin{figure}[ht]
\centering
\renewcommand{\thesubfigure}{\thefigure.\arabic{subfigure}}
\subfigure[]{\includegraphics[width=2.5in,height=1.3in]{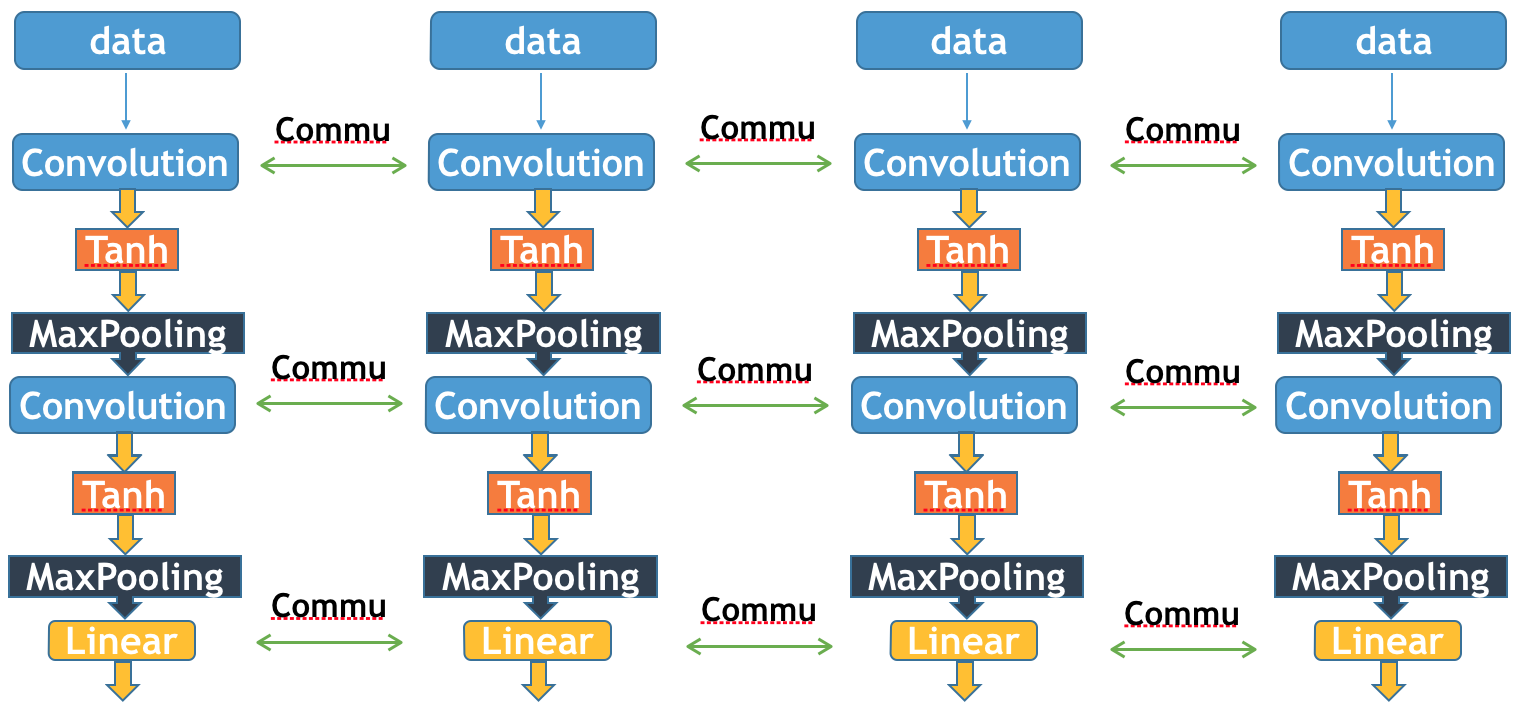}
\label{fig:data_parallelism}}
\subfigure[]{\includegraphics[width=0.6in,height=1.3in]{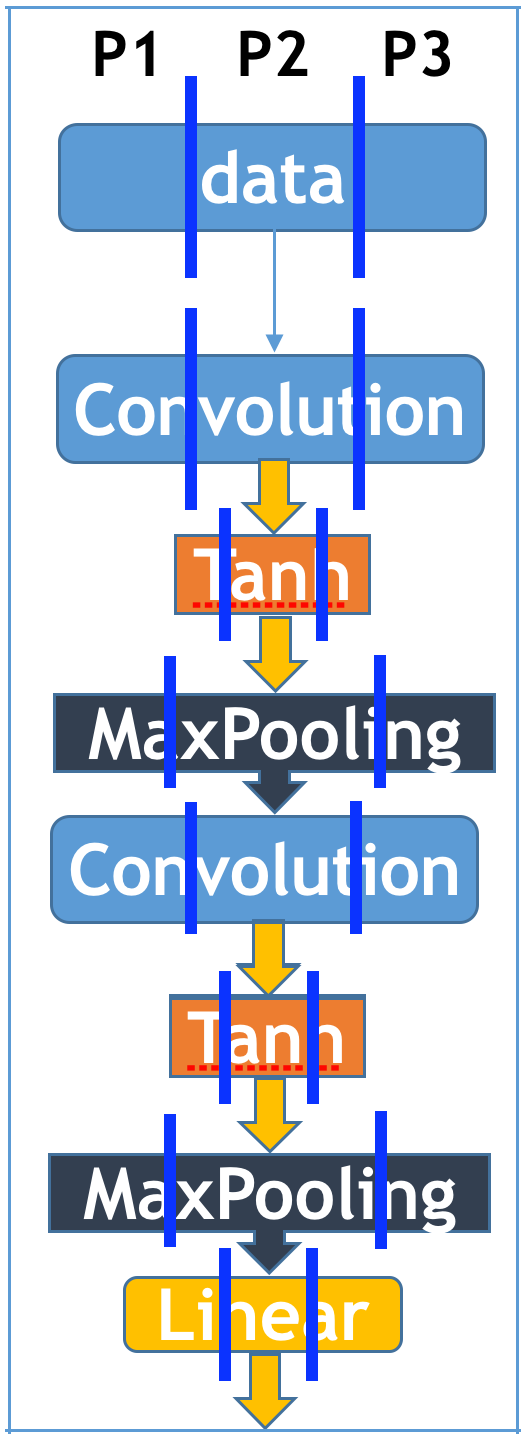}
\label{fig:model_parallelism}}
\caption{Methods of parallelism. \ref{fig:two_parallelism}.1 is data parallelism on 4 machines. \ref{fig:two_parallelism}.2 is model parallelism on 3 machines. Model parallelism partitions the neural network into $P$ pieces whereas data parallelism replicates the neural network itself in each processor.}
\label{fig:two_parallelism}
\end{figure}

Let us parallelize the DNN training process on $P$ machines. There are two major parallelism strategies for this: Data Parallelism (Fig. \ref{fig:two_parallelism}.1) and Model Parallelism (Fig. \ref{fig:two_parallelism}.2). All the later parallel methods are the variants of these two methods.

{\bf Data Parallelism} \cite{dean2012large} The dataset is partitioned into $P$ parts and each machine only gets one part. Each machine has a copy of the neural network, hence the weights ($W$). 
The communication includes sum of all the gradients $\Delta W_i$ and broadcast of $W$.
The first part of communication is conducted between Backward Propagation and Weights Update. The master updates $W$ by $W$ $\leftarrow$ $W -  \eta \sum_{i=1}^{P} \Delta W_i$ after it gets all the sub-gradients $\Delta W_i$ from the workers. Then the master machine broadcasts $W$ to all the worker machines, which is the second part of communication.
Figure \ref{fig:two_parallelism}.1 is an example of data parallelism on 4 machines.


{\bf Model Parallelism} \cite{coates2013deep} Data parallelism replicates the neural network itself on each machine while model parallelism partitions the neural network into $P$ pieces. Partitioning the neural network means parallelizing the matrix operations on the partitioned network. Thus, model parallelism can get the same solution as the single-machine case. 
Figure \ref{fig:two_parallelism}.2 shows model parallelism on 3 machines. These three machines partition the matrix operation of each layer.
However, because both the batch size ($<= 2048$) and the picture size (e.g. $32 \times 32$) typically are relatively small, the matrix operations are not large. For example, parallelizing a 2048$\times$1024$\times$1024 matrix multiplication only needs one or two machines. 
Thus, state-of-the-art methods often use data-parallelism (\cite{amodei2015deep}, \cite{chen2016revisiting}, \cite{dean2012large}, \cite{seide20141}).

\subsection{Evaluating Our Method}
The objective of this paper is designing distributed algorithms on HPC systems to get the same or higher classification accuracy (algorithm benchmark) in a shorter time. If our optimization may influence the convergence of algorithm, we report both the time and accuracy. Otherwise, we only report the time for experimental results. All algorithmic comparisons in this paper used the same hardware (e.g. \# CPUs, \# GPUs, and \# KNLs) and the same hyper-parameters (e.g. batch size, learning rate). We do not compare different architectures (e.g. KNL vs K80 GPU) because they have different performance, power, and prices.

\section{Related Work}
In this section, we review the previous literature about scaling deep neural networks on parallel or distributed systems.

\subsection{Parameter Server (Async SGD)}
\begin{figure}[!t]
\centering
\includegraphics[width=2.1in]{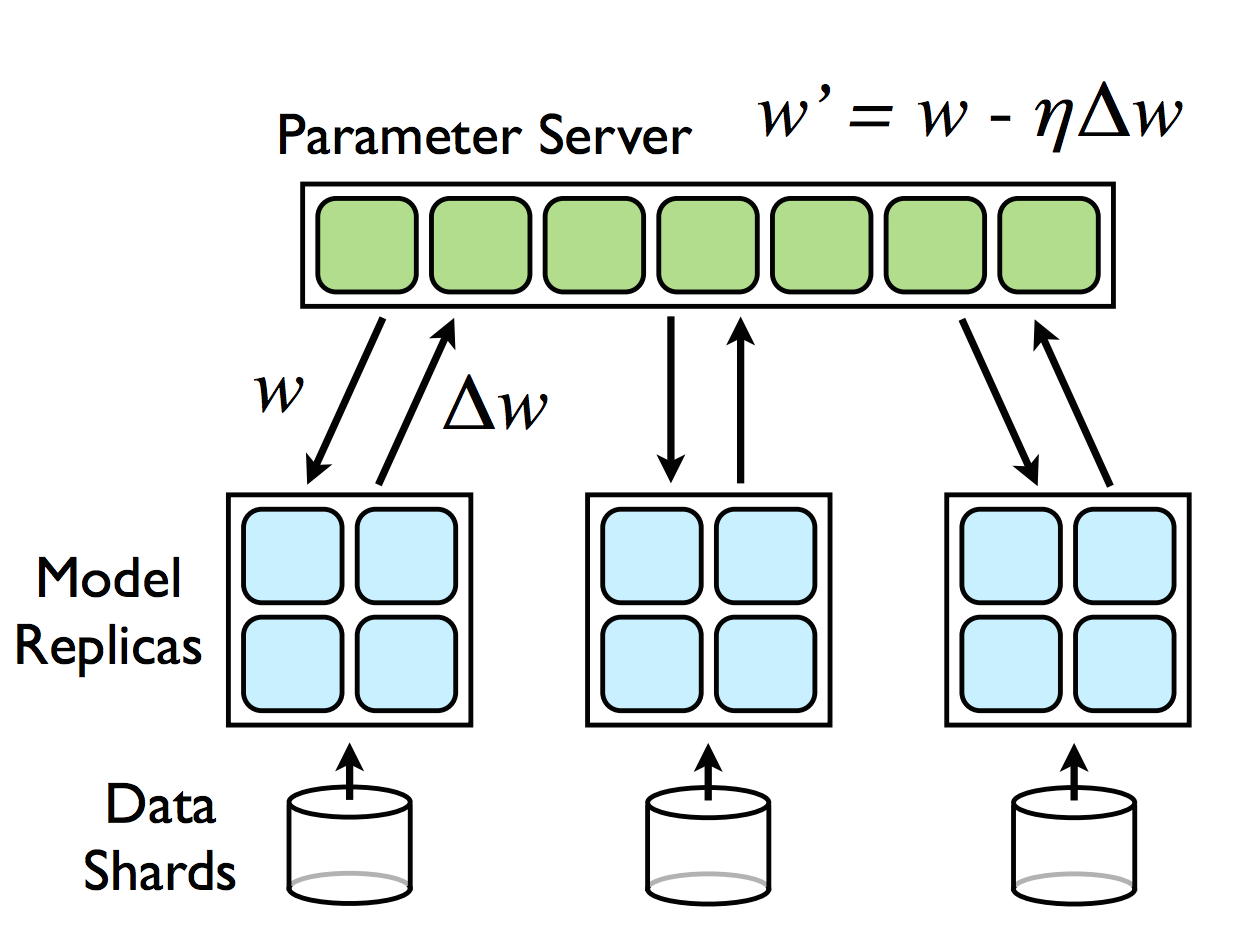}
\caption{This figure \cite{dean2012large} is an illustration of Parameter Server. Data is partitioned to workers. Each worker computes a gradient and sends it to server for updating the weight. The updated model is copied back to wokers}
\label{fig:ps}
\end{figure}

Figure \ref{fig:ps} illustrates the idea of parameter server or Asynchronous SGD \cite{dean2012large}. Under this framework, each worker machine has a copy of weight $W$. The dataset is partitioned to all the worker machines. At each step, $i$-th worker computes a sub-gradient ($\Delta W_i$) from its own data and weight. Then the $i$-th worker sends $\Delta W_i$ to the master ($i$ $\in$ $\{1, 2, ..., P\}$). The master receives $\Delta W_i$, conducts the weight update, and sends weight back to $i$-th worker machine. All the workers finish this step asynchronously, using first come first serve (FCFS).

\subsection{Hogwild (Lock Free)}
The Hogwild method \cite{recht2011hogwild} can be presented as a variant of Async SGD. The master machine is a shared memory system. For Async SGD, if the sub-gradient from $j$-th worker arrives during the period that the master is interacting with $i$-th worker, then $W$ $\leftarrow$ $W -  \eta \Delta W_j$ can not be started before $W$ $\leftarrow$ $W -  \eta \Delta W_i$ is finished ($i$, $j$ $\in$ $\{1, 2, ..., P\}$). This means that there is a lock to avoid weight update conflicts on the shared memory system (master machine). The lock makes sure the master only processes one sub-gradient at one time. The Hogwild method, however, removes the lock and allows the master to process multiple sub-gradients at the same time. The proof of Hogwild's lock-free convergence is in its paper \cite{recht2011hogwild}.

\subsection{EASGD (Round-Robin)\label{sec:easgd}}
The Elastic Averaging SGD (EASGD) method \cite{zhang2015deep} can also be presented as a variant of Async SGD. Async SGD uses a FCFS strategy for processing the sub-gradients asynchronously. EASGD uses a round-robin strategy for ordered update, i.e. $W$ $\leftarrow$ $W -  \eta \Delta W_i$ can not be started before $W$ $\leftarrow$ $W -  \eta \Delta W_{i-1}$ is finished ($i$ $\in$ $\{2, 3, ..., n\}$). Also, EASGD requires the workers to conduct the update locally (Equation (\ref{eq:global_local})). Before all the workers conduct the local updating, the master updates the center (or global) weight (Equation (\ref{eq:global_center})). The $\rho$ in Equation (\ref{eq:global_local}) and Equation (\ref{eq:global_center}) is a term that connects the global and local parameters. The framework of Original EASGD method is shown in Algorithm \ref{algo:original_easgd}.

\begin{equation}
  W_{t+1}^i = W_t^i - \eta({\Delta}W_t^i+\rho(W_t^i-\bar{W}_t))
  \label{eq:global_local}
\end{equation}

\begin{equation}
  \bar{W}_{t+1} = \bar{W}_t + \eta \sum_{i=1}^P \rho(W_t^i-\bar{W}_t)
  \label{eq:global_center}
\end{equation}

\begin{algorithm}
\DontPrintSemicolon 
\KwIn{samples and labels: $\{X_i, y_i\}$ $i \in {1, ..., n}$ \newline \#iterations: $T$, batch size: $b$, \#GPUs: $G$ }
\KwOut{model weight $W$}
Normalize $X$ on CPU by standard deviation: $E(X) = 0$ (mean) and $\sigma(X) = 1$ (variance)\;
Initialize $W$ on CPU: random and Xavier weight filling\;
\For{$j=1$; $j <= G$; j++} {
    create {\bf local} weight $W_j$ on $j$-th GPU, copy $W$ to $W_j$\;
}
create {\bf global} weight $\bar{W}_1$ on 0-th GPU, copy $W$ to $\bar{W}_1$\;
\For{$t=1$; $t <= T$; t++} {
    j = t mod G\;
    CPU {\bf randomly} picks $b$ samples\; 
    CPU {\bf asynchronously} copies $b$ samples to $j$-th GPU\;
    CPU sends $\bar{W}_t$ to $j$-th GPU\;
    Forward and Backward Propagation on $j$-th GPU\;
    CPU gets $W_t^j$ from $j$-th GPU\;
    $j$-th GPU updates $W_t^j$ by Equation (\ref{eq:global_local})\;
    CPU updates $\bar{W}_t$ by $\bar{W}_{t+1} = \bar{W}_t + \eta \rho(W_t^j-\bar{W}_t)$\;
}
\caption{Original EASGD on Multi-GPU system\;master: CPU, workers: GPU$_1$, GPU$_2$, ..., GPU$_P$}
\label{algo:original_easgd}
\end{algorithm}

\subsection{Other methods}
Li et al. \cite{li2016optimizing} is focused on single-node memory optimization. The idea is included in our implemenation. There is some work \cite{coates2013deep}, \cite{le2013building} on scaling up deep neural networks by model parallelism method, which is out of scope for this paper. This paper is focused on data parallelism.
Low-precision representation of neural networks is another direction of research. The idea is to use low-precision floating point to reduce the computation and communication for getting the acceptable accuracy (\cite{courbariaux2014training}, \cite{gupta2015deep}, \cite{hubara2016quantized}, \cite{seide20141}). We reserve this for future study. 

\section{Experimental Setup}
\subsection{Experimental Datasets}
Our test datasets are Mnist \cite{lecun1998gradient}, Cifar \cite{krizhevsky2009learning}, and ImageNet \cite{deng2009imagenet}, which are the standard benchmarks for deep learning research. Descriptions can be found in Table \ref{tab:dataset}. The application of Mnist dataset is hardwritten digits recognition. The images of Mnist were grouped into 10 classes (0, 1, 2, ..., 9). The application of Cifar dataset is object recognition. Cifar dataset includes 10 classes: airplane, automobile, bird, cat, deer, dog, frog, horse, ship, truck. Each Cifar image only belongs to one class. The accuracy of random guess for Mnist and Cifar image prediction is 0.1.

ImageNet \cite{deng2009imagenet} is a computer vision dataset of over 15 million labeled images belonging to more than 20,000 classes. The images were collected from the web and labeled by human labelers using Amazon's Mechanical Turk crowd-sourcing tool. An annual competition called the ImageNet Large-Scale Visual Recognition Challenge (ILSVRC) has been held since 2010. ILSVRC uses a subset of ImageNet with 1200 images in each of 1000 classes. In all, there are roughly 1.2 million training images, 50,000 validation images, and 150,000 testing images. In this paper, the ImageNet dataset means ILSVRC-2012 dataset.
The accuracy of random guess for ImageNet image prediction is 0.001.

\begin{table}
\small
  \caption{The Test Datasets}
  \label{tab:dataset}
  \begin{tabular}{l*{9}{c}r}
    \toprule
    Dataset & Training Images & Test Images & Pixels & Classes\\
    \midrule
    Mnist \cite{lecun1998gradient} & 60,000 & 10,000 & 28$\times$28 & 10\\
    Cifar \cite{krizhevsky2009learning} & 50,000 & 10,000 & 3$\times$32$\times$32 & 10\\
    ImageNet \cite{deng2009imagenet} & 1.2 million & 150,000 & 256$\times$256 & 1000\\
  \bottomrule
\end{tabular}
\end{table}

\subsection{Neural Network Models\label{sec:model}}
We use the state-of-the-art DNN models to process the datasets in this paper.
The Mnist dataset was processed by LetNet \cite{lecun1998gradient}, which is shown in Figure \ref{fig:cnn}. The Cifar dataset is processed by AlexNet \cite{krizhevsky2012imagenet}, which has 5 convolutional layers and 3 fully-connected layers. ImageNet dataset is processed by GoogleNet \cite{szegedy2015going} and VGG \cite{szegedy2015going}. GoogleNet has 22 layers and VGG has 19 layers.

\subsection{The baseline}
In Section \ref{sec:distributed_algo}, the Original EASGD is our baseline. The original EASGD method (Algorithm \ref{algo:original_easgd}) uses round-robin approach for scheduling the way the master interacts with the workers. At any moment, the master can interact with just a single worker. Additionally, the interactions of different workers are ordered. The ($i+1$)-st worker can not begin before $i$-th finishes. 

\section{Distributed Algorithm Design} \label{sec:distributed_algo}

\subsection{Redesigning the parallel SGD methods}
In this section we redesign some efficient parallel SGD methods based on the existing methods (i.e. Original EASGD, Async SGD, Async MSGD, and Hogwild SGD). We will use our methods to make comparisons with the existing methods (i.e. we will plot accuracy versus time, on the same data sets and computing resources). Since the existing SGD methods were originally implemented on GPUs, we also implement our methods on GPUs. These ideas work in the same way for KNL chips because these methods are focused on inter-chip processing rather than intra-chip processing.

\begin{figure}[ht]
\centering
\renewcommand{\thesubfigure}{\thefigure.\arabic{subfigure}}
\subfigure[]{\includegraphics[width=1.6in]{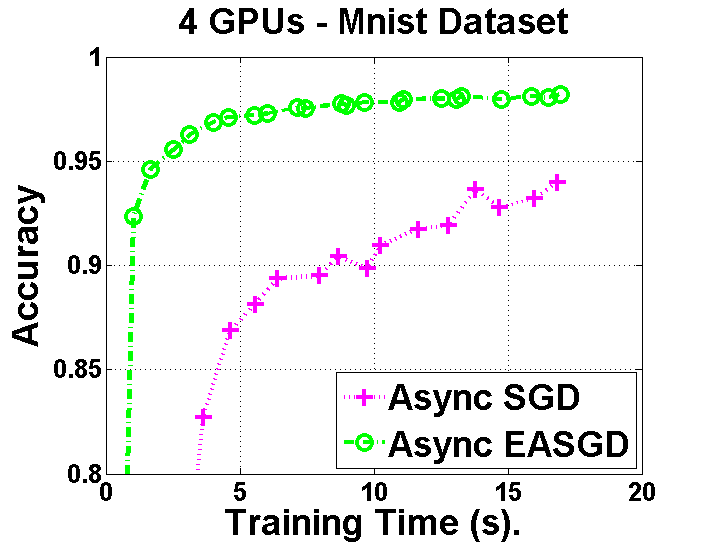}
\label{gpu_async}}
\subfigure[]{\includegraphics[width=1.6in]{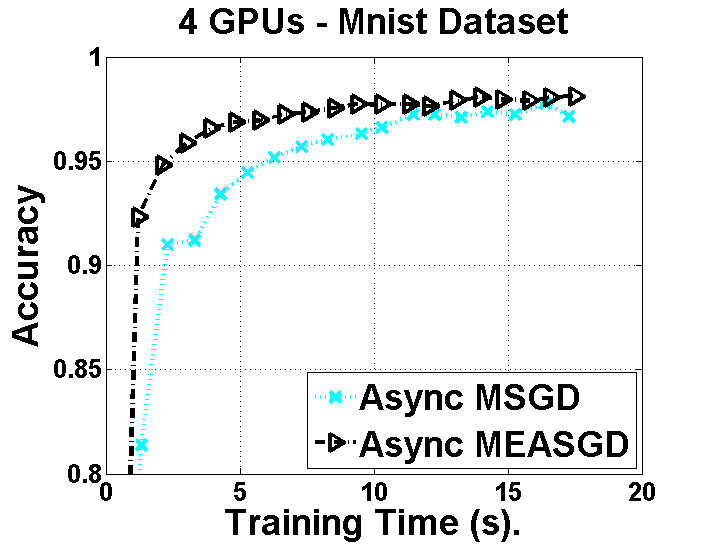}
\label{fig_second_case}}
\subfigure[]{\includegraphics[width=1.6in]{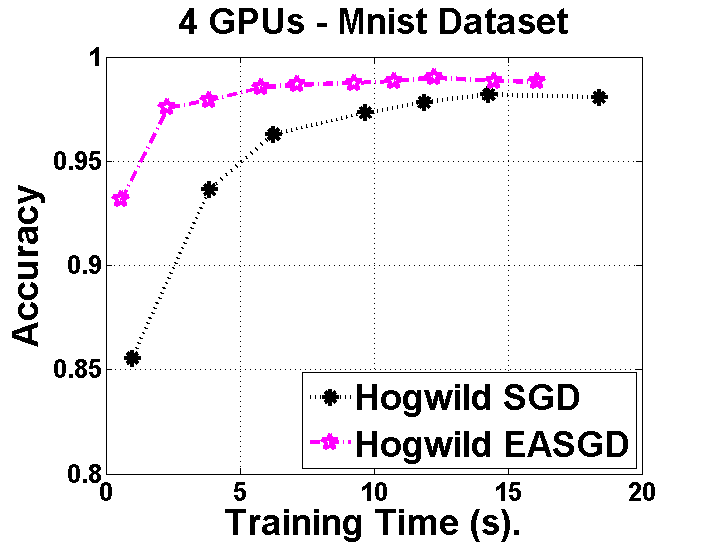}
\label{fig_third_case}}
\subfigure[]{\includegraphics[width=1.6in]{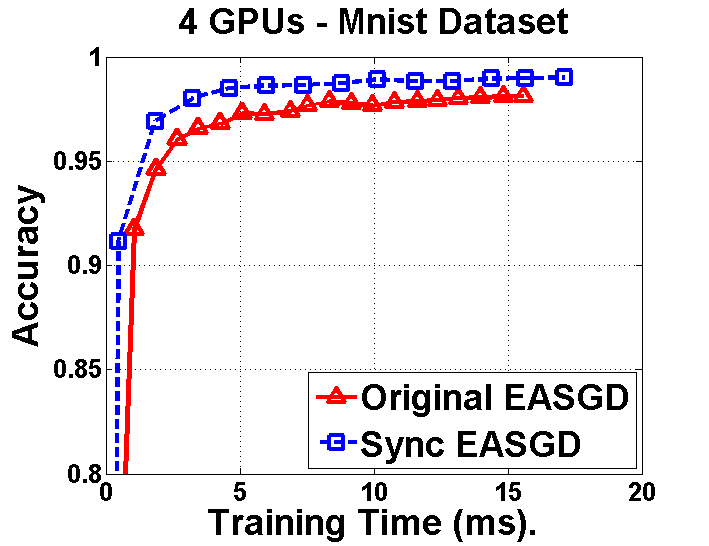}
\label{fig_fourth_case}}
\caption{Original EASGD, Hogwild SGD, Async SGD, and Async MSGD are the existing methods. All the comparisons use the same hardware and data. Our methods are faster. Each point on the figure is a single train and test. For example, Hogwild EASGD has 10 points in the figure. It means we run 10 mutually independent Hogwild EASGD cases with different numbers of iterations (e.g. 1k, 2k, 3k, ..., 10k). \textcolor{black}{The experiments are conducted on 4 Tesla M100 GPUs that are connected with a 96-lane, 6-way PCIe switch.}}
\label{fig:p2p_compare}
\end{figure}

{\bf Async EASGD} The original EASGD method (Algorithm \ref{algo:original_easgd}) uses round-robin approach for scheduling. This method is inefficient because the computation and update of different GPUs are ordered (Section \ref{sec:easgd}). The ($i+1$)-st worker can not begin before $i$-th finishes. Although this method has good fault-tolerance and convergence properties, it is inefficient. Therefore, our first optimization is to use parameter-server update to replace the round-robin update. \textcolor{black}{The difference between our Async EASGD and Original EASGD is that we use first-come first-served (FCFS) strategy to process multiple workers while they use ordered rule to process multiple workers.} We put the global (or center) weight $\bar{W}$ on the master machine. The $i$-th worker machine has its own local weight $W^i$. During the $t$-th iteration, there are three steps: 

\begin{itemize}
\item (1) $i$-th worker first sends its local weight $W^i_t$ to master and master returns $\bar{W}_t$ to $i$-th worker ($i \in \{1, 2, ..., P\}$). 

\item (2) $i$-th worker computes gradient $\Delta W_t^i$ and receives $\bar{W}_t$. 

\item (3) master does the update based on Equation (\ref{eq:global_center}) and worker does the update based on Equation (\ref{eq:global_local}). 
\end{itemize}

From Figure \ref{fig:p2p_compare}.1 we can observe that our method Async EASGD is faster than Async SGD.

{\bf Async MEASGD} Momentum \cite{sutskever2013importance} is an important method to accelerate SGD. The updating rule of Momentum SGD (MSGD) is shown in Equations (\ref{eq:momentum1}) and (\ref{eq:momentum2}). $V$ is the momentum parameter, which has the same dimension as the weight and gradient. $\mu$ is the momentum rate. Rule of thumb is $\mu$ = 0.9 or a similar value. In our design, the updating rule of MEASGD master will be the same as before (Equation (\ref{eq:global_center})). The updating rule of the $i$-th worker will be changed to Equations (\ref{eq:easgdmomentum1}) and (\ref{eq:easgdmomentum2}). From Figure \ref{fig:p2p_compare}.2 we can observe that our method Async MEASGD is faster and more stable than Async MSGD.

\begin{equation}
  V_{t+1} = \mu V_t - \eta{\Delta}W_t
  \label{eq:momentum1}
\end{equation}

\begin{equation}
  W_{t+1} = W_{t} + V_{t+1}
  \label{eq:momentum2}
\end{equation}

\begin{equation}
  V_{t+1}^i = \mu V_t^i - \eta{\Delta}W_t^i
  \label{eq:easgdmomentum1}
\end{equation}

\begin{equation}
  W_{t+1}^i = W_{t}^i + V_{t+1}^i - \eta \rho(W_t^i-\bar{W}_t)
  \label{eq:easgdmomentum2}
\end{equation}

{\bf Hogwild EASGD} For Hogwild SGD, the lock for updating W is removed to achieve a faster convergence. In the same way, for regular EASGD, there should be a lock between $\bar{W}_{t+1} = \bar{W}_t + \eta \rho(W_t^i-\bar{W}_t)$ and $\bar{W}_{t+1} = \bar{W}_t + \eta \rho(W_t^j-\bar{W}_t)$ ($i, j \in \{1, 2, ..., P\}$). The reason is that $W_t^i$ and $W_t^j$ may arrive at the same time. Thus, we remove this lock to get the Hogwild EASGD method. From Figure \ref{fig:p2p_compare}.3 we clearly observe that Hogwild EASGD is much faster than Hogwild SGD. The convergence proof of Hogwild EASGD can be found in the appendix\footnote{\href{https://www.cs.berkeley.edu/~youyang/HogwildEasgdProof.pdf}{https://www.cs.berkeley.edu/$\sim$youyang/HogwildEasgdProof.pdf}}.

{\bf Sync EASGD} The updating rules of Sync EASGD are Equations (\ref{eq:global_local}) and (\ref{eq:global_center}). The Sync EASGD contains five steps at iteration $t$: 

\begin{itemize}
\item (1) the $i$-th worker computes its sub-gradient $\Delta W_t^i$ based on its data and weight $W_t^i$ ($i \in \{1, 2, ..., P\}$). 

\item (2) the master broadcasts $\bar{W_t}$ to all the workers. 

\item (3) the system does a reduce operation to get ${\sum}_{i=1}^{P} W_t^i$ and sends it to master. 

\item (4) the $i$-th worker updates its local weight $W_t^i$ based on Equation (\ref{eq:global_local}). 

\item (5) the master updates $\bar{W_t}$ based on Equation (\ref{eq:global_center}). 
\end{itemize}

Among them, step (1) and step (2) can be overlapped, step (4) and step (5) can be overlapped. From Figure \ref{fig:p2p_compare}.4 we observe that Sync EASGD is faster than Original EASGD. Here, Sync EASGD means Sync EASGD3 implementation detailed in Section \ref{sec:multigpu}.

\begin{figure}[ht]
\centering
\renewcommand{\thesubfigure}{\thefigure.\arabic{subfigure}}
\subfigure[Sync EASGD1]{\includegraphics[width=3.0in]{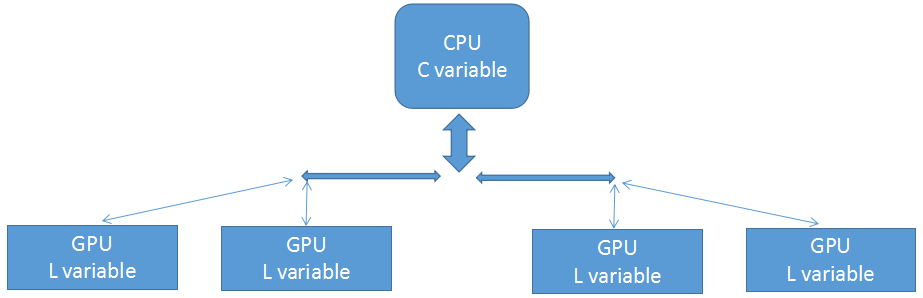}
\label{fig_first_case}}
\subfigure[Sync EASGD2 and Sync EASGD3]{\includegraphics[width=3.0in]{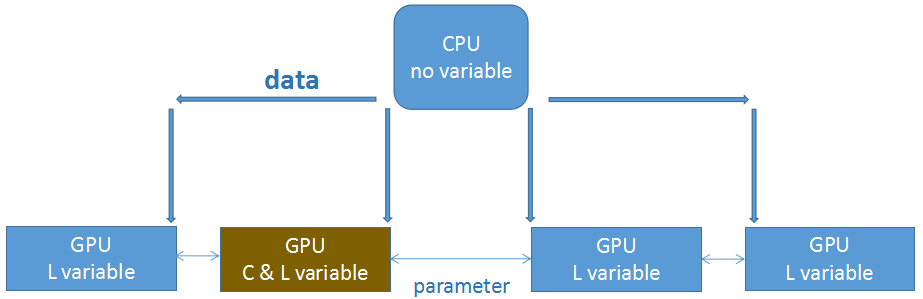}
\label{fig_first_case}}
\caption{The architecture our Sync EASGD design on multi-GPU system. C variable means Center Weight or Global Weight, L variable means Local Weight.}
\label{fig:sync_easgd_architecture}
\end{figure}

\begin{figure}[!t]
\centering
\includegraphics[width=3.8in]{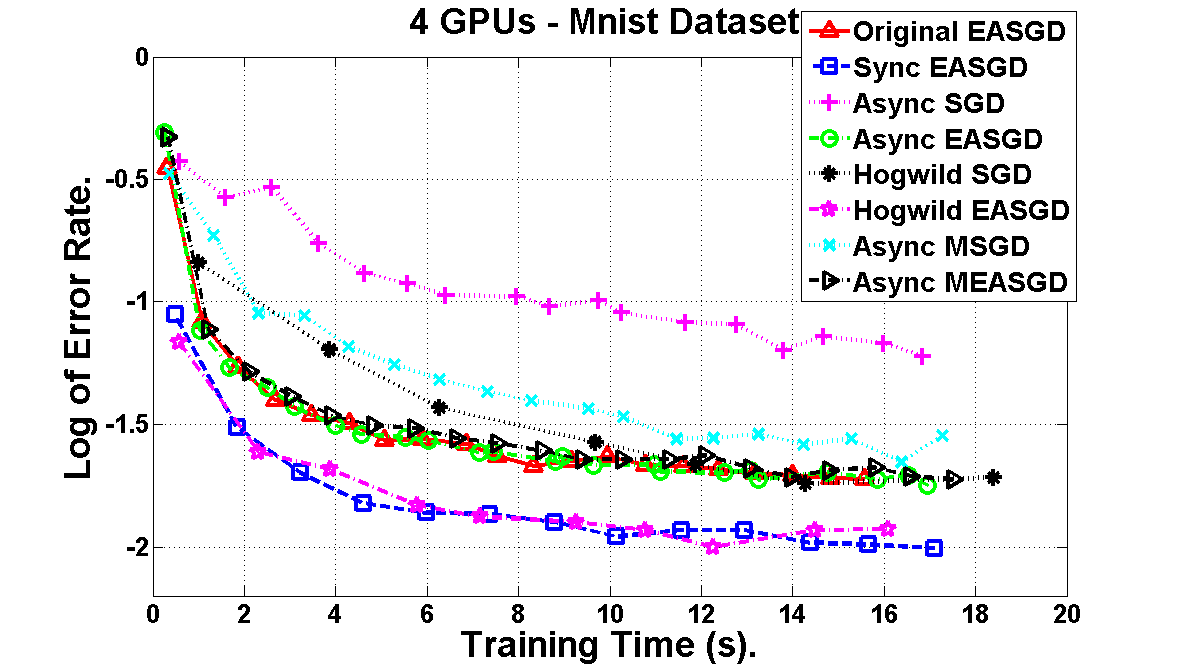}
\caption{To visualize the comparisons, we use error rate ($1.0$ $-$ accuracy) as the algorithm benchmark. Then we use $log_{10}$ scale of error rate to make the comparisons more clear.  Among these methods, Original EASGD, Hogwild SGD, Async SGD, and Async MSGD are the existing methods. The rest of them are our methods. Each point on the figure is a single run. For example, Sync EASGD has 13 points in the figure. It means we run 13 mutually independent Sync EASGD cases with different numbers of iterations.
It also means longer time or more iterations will help us to get a higher accuracy, even with different initiations.
\textcolor{black}{The experiments are conducted on 4 Tesla M100 GPUs that are connected with a 96-lane, 6-way PCIe switch.}}
\label{fig:compare}
\end{figure}

\begin{figure}[!t]
\centering
\includegraphics[width=3.2in]{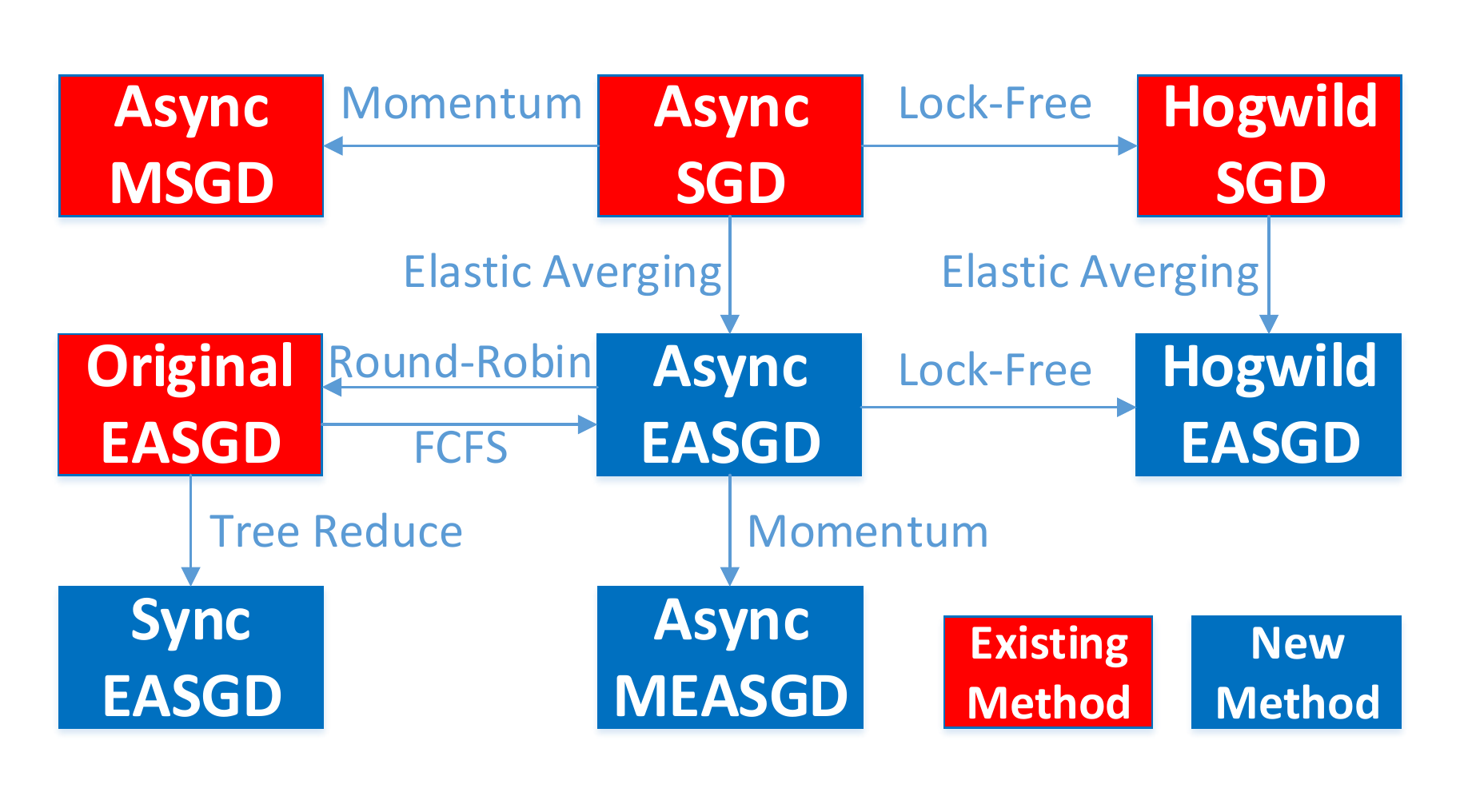}
\caption{This framework of our algorithm design. The red block means the existing method and the blue block means the new method.}
\label{fig:framework}
\end{figure}

We make an overall comparison by putting these comparisons together into Figure \ref{fig:compare}. Among them, Original SGD, Hogwild SGD, Async SGD, and Async MSGD are the existing methods. Our method is always faster than its counterpart as already seen in Figure \ref{fig:p2p_compare}. We also observe that Sync EASGD or Hogwild EASGD is the fastest method among them. Sync EASGD and Hogwild EASGD are essentially tied for fastest. Sync EAGSD incorporates a number of optimizations that we describe in more detail in sections \ref{sec:one_layer} and \ref{sec:algorithm_codesign}. The framework of our algorithm design is shown in Fig. \ref{fig:framework}, which shows the difference between these methods.

\begin{table}
  \caption{InfiniBand Performance under $\alpha$-$\beta$ Model}
  \label{tab:alphabeta}
  \begin{tabular}{l*{9}{c}r}
    \toprule
    Network & $\alpha$ (latency) & $\beta$ (1/bandwidth)\\
    \midrule
    Mellanox 56Gb/s FDR IB & $0.7\times10^{-6}$s & $0.2\times10^{-9}$s\\
    Intel 40Gb/s QDR IB & $1.2\times10^{-6}$s & $0.3\times10^{-9}$s\\
    Intel 10GbE NetEffect NE020 & $7.2\times10^{-6}$s & $0.9\times10^{-9}$s\\
      \bottomrule
\end{tabular}
\end{table} 

\subsection{Single-Layer Communication} \label{sec:one_layer}
\begin{figure}[!t]
\centering
\includegraphics[width=2.3in]{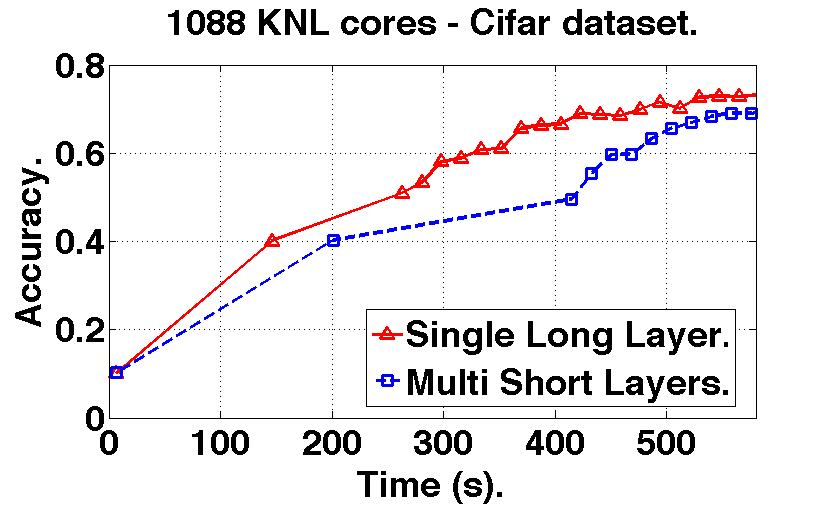}
\caption{The benefit of packed layer comes from reduced communication latency and continuous memory access. Since this is Sync SGD, the red triangles and blue squares should be at identical heights. The reason for different heights is that a different random number generator seed is used for the two runs. The example used Sync SGD to process AlexNet (Section \ref{sec:model}).}
\label{fig:singleLayer}
\end{figure}

Current deep learning systems \cite{iandola2016firecaffe} allocate noncontiguous memory for different layers of the neural networks. They also conduct multiple rounds of communication for different layers. We allocate the neural networks in a contiguous way and pack all the layers together and conduct one communication each time. This significantly reduces the latency. From Figure \ref{fig:singleLayer} we can observe the benefit of this technique. There are two reasons for the improvement: (1) The communication overhead of sending a n-word message can be formulated as $\alpha$-$\beta$ model: ($\alpha$ + $\beta \times n$) seconds. $\alpha$ is the network latency and $\beta$ is the reciprocal of network bandwidth. $\beta$ is much smaller than $\alpha$, which is the major communication overhead (Table \ref{tab:alphabeta}). Thus, for transferring the same volume of data, sending one big message is better than multiple small messages. (2) The continuous memory access has a higher cache-hit ratio than the non-continuous memory access.

\begin{algorithm}
\DontPrintSemicolon 
\KwIn{samples and labels: $\{X_i, y_i\}$ $i \in {1, ..., n}$ \newline \#iterations: $T$, batch size: $b$, \# GPUs: $G$ }
\KwOut{model weight $W$}
Normalize $X$ on CPU by standard deviation: $E(X) = 0$ (mean) and $\sigma(X) = 1$ (variance)\;
Initialize $W$ on CPU: random and Xavier weight filling\;
\For{$j=1$; $j <= G$; j++} {
    create {\bf local} weight $W_1^j$ on GPU$_j$, copy $W$ to $W_1^j$\;
   }
create {\bf global} weight $\bar{W}_1$ on CPU, copy $W$ to $\bar{W}_1$\;
\For{$t=1$; $t <= T$; t++} {
    \For{$j=1$; $j <= G$; j++} {
    	CPU {\bf randomly} picks $b$ samples\;
    	CPU {\bf asynchronously} copies $b$ samples to $j$-th GPU$_j$\;
    }
    	Forward and Backward Propagation on all the GPUs\;
	CPU broadcasts $\bar{W}_t$ to all the GPUs\;
	CPU gets ${\sum}_{j=1}^{G} W_t^j$ from all the GPUs\;
    	All the GPUs update $W_t^j$ by Equation (\ref{eq:global_local})\;
    	CPU updates $\bar{W}_t$ by Equation (\ref{eq:global_center})\;
    }
\caption{Sync EASGD1\;master: CPU, workers: GPU$_1$, GPU$_2$, ..., GPU$_P$}
\label{algo:sync_easgd1}
\end{algorithm}

\section{Algorithm-Architecture Codesign} \label{sec:algorithm_codesign}

\subsection{Multi-GPU Optimization} \label{sec:multigpu}
In this section we show how we optimize EASGD step-by-step on a multi-GPU system. We use {\bf Sync EASGD1}, {\bf Sync EASGD2}, and {\bf Sync EASGD3} to illustrate our three-step optimization.

\subsubsection{Sync EASGD1}
Algorithm \ref{algo:original_easgd} is the original EASGD algorithm. Multi-GPU system implementation contains 8 potentially time-consuming parts. For Algorithm \ref{algo:original_easgd}, they are: (1) data I/O (Input and Output); (2) data and weight initialization (lines 1-2); (3) GPU-GPU parameter communication (none); (4) CPU-GPU data communication (line 9); (5) CPU-GPU parameter communication (lines 10 and 12); (6) Forward and Backward propagation (line 11); (7) GPU weight update (line 13); (8) CPU weight update (line 14). We ignore parts (1) and (2) because they only cost a tiny percent of time. GPU-GPU parameter communication means different GPUs exchange weights. CPU-GPU data communication means GPU copies a batch of samples each iteration. CPU-GPU parameter communication means CPU sends global weight $\bar{W}$ to GPUs and receives local weights $W^i$ ($i \in \{1, 2, ..., P$) from GPUs.
Parts (3), (4), and (5) are communication. Parts (6), (7), and (8) are computation. After benchmarking the code, we found the major overhead of EASGD is communication (Figure \ref{fig:commu_compu}), which costs 87\% of the total training time on an 8-GPU system. If we look deep into the communication, we observe that CPU-GPU parameter communication costs much more time than CPU-GPU data communication (86\% vs 1\%). The reason is that the size of weights (number of elements in $W$) is much larger than a batch of training data. For example, the weights of AlexNet are 249 MB while 64 Cifar samples are only 64 $\times$ 32 $\times$ 32 $\times$ 3 $\times$ 4B = 768 KB. To solve this problem, we design Sync EASGD1 (Algorithm \ref{algo:sync_easgd1}). In Sync EASGD1, $P$ blocking send/receive operations can be efficiently processed by a tree-reduction operation \textcolor{black}{(e.g. standard MPI reduction)}, which reduces the communication overhead from $P(\alpha + |W| \beta)$ to $logP(\alpha + |W| \beta)$. Our experiments show that Sync EASGD1 achieves a $3.7\times$ speedup over Original EASGD (Table \ref{tab:commu_compu} and Figure \ref{fig:commu_compu}). 

\begin{algorithm}
\DontPrintSemicolon 
\KwIn{samples and labels: $\{X_i, y_i\}$ $i \in {1, ..., n}$ \newline \#iterations: $T$, batch size: $b$, \# GPUs: $G$ }
\KwOut{model weight $W$}
Normalize $X$ on CPU by standard deviation: $E(X) = 0$ (mean) and $\sigma(X) = 1$ (variance)\;
Initialize $W$ on CPU: random and Xavier weight filling\;
\For{$j=1$; $j <= G$; j++} {
    create {\bf local} weight $W_1^j$ on GPU$_j$, copy $W$ to $W_1^j$\;
   }
create {\bf global} weight $\bar{W}_1$ on GPU$_1$, copy $W$ to $\bar{W}_1$\;
\For{$t=1$; $t <= T$; t++} {
    \For{$j=1$; $j <= G$; j++} {
    	CPU {\bf randomly} pick $b$ samples\;
    	CPU {\bf asynchronously} copy $b$ samples to $j$-th GPU$_j$\;
    }
    	Forward and Backward Propagation on all the GPUs\;
	GPU$_1$ broadcasts $\bar{W}_t$ to all the GPUs\;
	GPU$_1$ gets ${\sum}_{j=1}^{G} W_t^j$ from all the GPUs\;
    	All the GPUs update $W_t^j$ by Equation (\ref{eq:global_local})\;
    	GPU$_1$ updates $\bar{W}_t$ by Equation (\ref{eq:global_center})\;
    }
\caption{Sync EASGD2 \& Sync EASGD3\;master: GPU$_1$, workers: GPU$_1$, GPU$_2$, ..., GPU$_P$}
\label{algo:sync_easgd_2_3}
\end{algorithm}

\subsubsection{Sync EASGD2}
From Table \ref{tab:commu_compu} we observe that CPU-GPU communication is still the major overhead of communication. Thus, we want to move either data or weights from CPU to GPU to reduce the communication overhead.
We can not put all the data on the GPU card because the on-chip memory is very limited compared with CPU. For example, the training part of ImageNet dataset is 240 GB while the on-chip memory of K80 is only around 12 GB. Since the algorithm needs to randomly pick samples from the dataset, we can not predict which part of dataset will be used by a certain GPU. Thus, we put all the training and test data on the CPU. We only copy the required data to the GPUs at runtime each iteration. On the other hand, the weights are usually smaller than 1 GB, which can be stored on a GPU card. For example, the large DNN model VGG-19 \cite{simonyan2014very} is 575 MB. Also, the weight will be reused every iteration (Algorithm \ref{algo:sync_easgd_2_3}). Thus, we put all the weights on GPU to reduce communication overhead. We refer to this method as Sync EASGD2, which achieves 1.3$\times$ speedup over Sync EASGD1. The framework of Sync EASGD2 is shown in Algorithm \ref{algo:sync_easgd_2_3}.

\subsubsection{Sync EASGD3}
\begin{figure}[ht]
\centering
\renewcommand{\thesubfigure}{\thefigure.\arabic{subfigure}}
\subfigure[]{\includegraphics[width=3.7in]{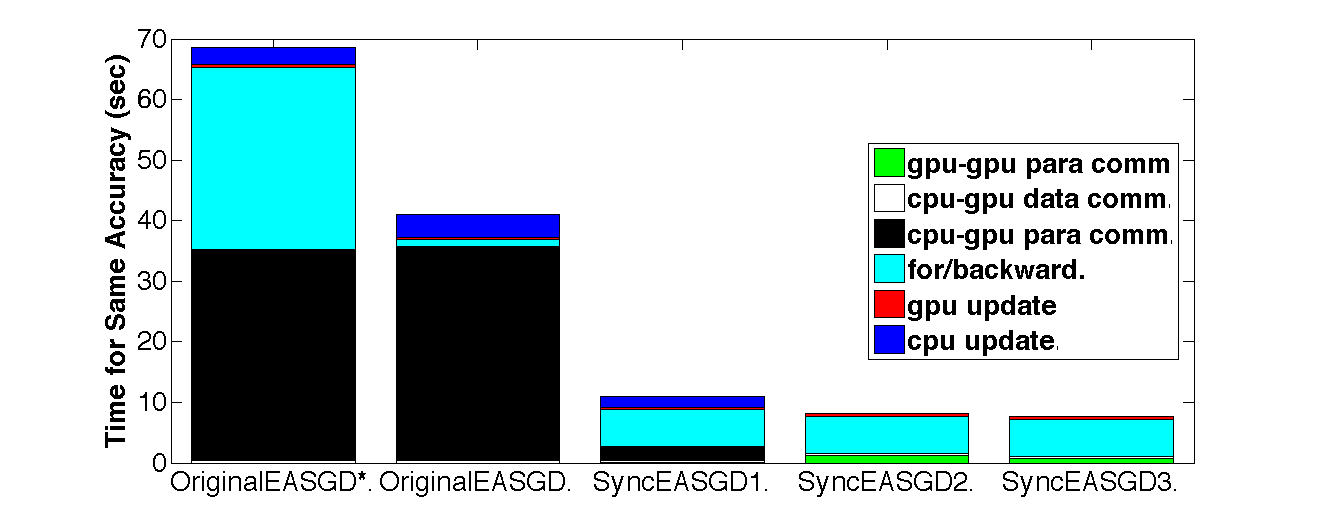}
\label{fig_first_case}}
\subfigure[]{\includegraphics[width=3.7in]{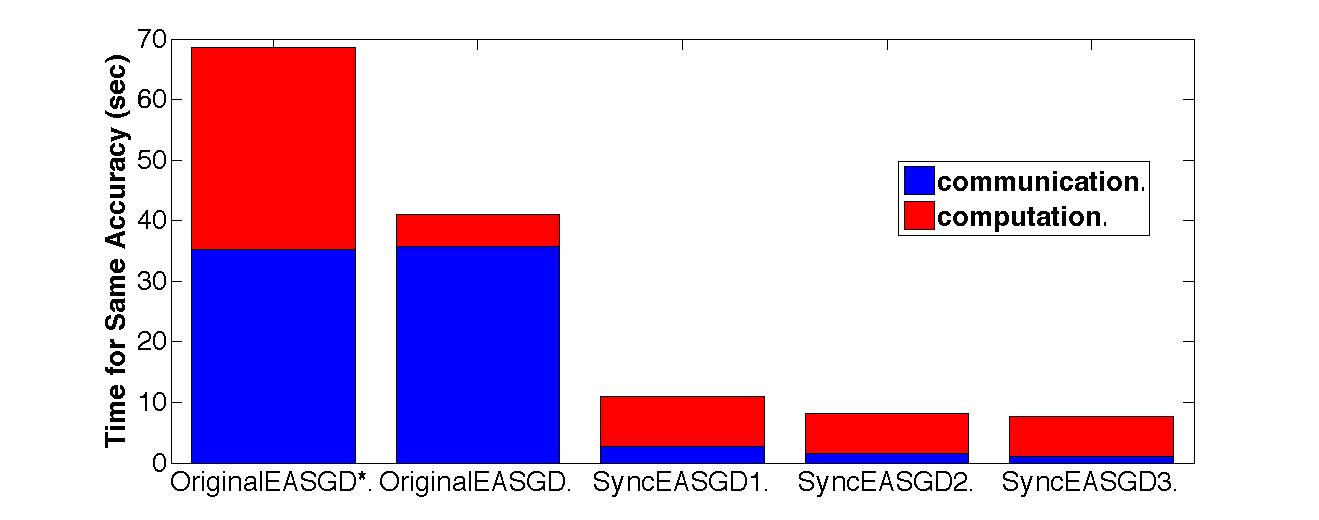}
\label{fig_first_case}}
\caption{Breakdown of time for EASGD variants. {\bf para} means parameter or weight. {\bf comm} means communication. Computation includes forward/backward, gpu update, and cpu update. \textcolor{black}{There is an overlap between for/backward and cpu-gpu para comm for Original EASGD. Let us refer to the non-overlap version as Original EASGD*.} Sync EASGD3 reduced the communication percentage from 87\% to 14\% and got 5.3$\times$ speedup over original EASGD for the same accuracy (98.8\%). The test is for MNIST dataset on 4 GPUs. More information is in Table \ref{tab:commu_compu}.}
\label{fig:commu_compu}
\end{figure}

We further improve the algorithm by overlapping the computation with the communication. We maximize the overlapping benefit inside the steps 7-14 of Algorithm \ref{algo:sync_easgd_2_3}. Because Forward/Backward Propagation uses the data from the CPU, steps 7-10 are a critical path. The GPU-GPU communication (steps 11-12) is not dependent on steps 7-10. Thus, we overlap steps 7-10 and steps 11-12 in Algorithm \ref{algo:sync_easgd_2_3}, yielding Sync EASGD3, which achieves a 1.1$\times$ speedup over Sync EASGD2. In all, Sync EASGD3 reduced the communication ratio from 87\% to 14\% and achieves 5.3$\times$ speedup over original EASGD for getting the same accuracy (Table \ref{tab:commu_compu} and Figure \ref{fig:commu_compu}). Thus we refer to Sync EASGD3 as {\bf Communication Efficient EASGD}. We also design similar algorithm for KNL cluster, which is shown in Algorithm \ref{algo:sync_easgd_knl}, discussed next.

\begin{algorithm}
\DontPrintSemicolon 
\KwIn{samples and labels: $\{X_i, y_i\}$ $i \in {1, ..., n}$ \newline \#iterations: $T$, batch size: $b$, \# KNL Nodes: $K$ }
\KwOut{model weight $W$}
All nodes read the samples and labels from disk\;
Normalize $X$ on all KNLs by standard deviation: $E(X) = 0$ (mean) and $\sigma(X) = 1$ (variance)\;
Initialize $W$ on 1-st KNL: random and Xavier weight filling\;
KNL$_1$ broadcasts $W$ to all KNLs\;
\For{$j=1$; $j <= K$; j++ {\bf parallel}} {
    create {\bf local} weight $W_1^j$ on KNL$_j$, copy $W$ to $W_1^j$\;
   }
create {\bf global} weight $\bar{W}_1$ on KNL$_1$, copy $W$ to $\bar{W}_1$\;
\For{$t=1$; $t <= T$; t++} {
    \For{$j=1$; $j <= K$; j++ {\bf parallel}} {
    	KNL$_j$ {\bf randomly} pick $b$ samples from local memory\;
    }
    	Forward and Backward Propagation on all the KNLs\;
	KNL$_1$ broadcasts $\bar{W}_t$ to all the KNLs\;
	KNL$_1$ gets ${\sum}_{j=1}^{K} W_t^j$ from all the KNLs\;
    	All the KNLs update $W_t^j$ by Equation (\ref{eq:global_local})\;
    	KNL$_1$ updates $\bar{W}_t$ by Equation (\ref{eq:global_center})\;
    }
\caption{Communication Efficient EASGD on KNL cluster master: KNL$_1$, workers: KNL$_1$, KNL$_2$, ..., KNL$_P$}
\label{algo:sync_easgd_knl}
\end{algorithm}

\begin{table*}
\small
  \caption{Breakdown of time for EASGD variants. {\bf para} means parameter or weight. {\bf comm} means communication. Computation includes for/backward, gpu update, and cpu update. Sync EASGD3 reduced the communication percentage from 87\% to 14\% and got 5.3$\times$ speedup over original EASGD for the same accuracy (98.8\%). There is an overlap between for/backward and cpu-gpu para comm for Original EASGD. Let us refer to the non-overlap version as Original EASGD*. The reason why Original EASGD*'s for/backward time is larger than the Sync methods (30s vs 6s) is that only one GPU (worker) is working at each iteration. Original EASGD/EASGD* need more iterations to get the same accuracy. The test is for Mnist dataset on 4 GPUs.}
  \label{tab:commu_compu}
  \begin{tabular}{l*{9}{c}r}
    \toprule
    Method & accuracy & iterations & time & gpu-gpu para & cpu-gpu data & cpu-gpu para & for/backward & gpu update & cpu update & comm ratio\\
    \midrule
    Original EASGD* & 0.988 & 5,000 & 69s & 0\% & 0.5\% & 51\% & 44\% & 0.5\% & 4\% & 52\%\\
    Original EASGD & 0.988 & 5,000 & 41s & 0\% & 1\% & 86\% & 3\% & 1\% & 9\% & 87\%\\
    Sync EASGD1 & 0.988 & 1,000 & 11s & 1\% & 3\% & 21\% & 55\% & 4\% & 16\% & 25\%\\
    Sync EASGD2 & 0.988 & 1,000 & 8.2s & 16\% & 4\% & 0\% & 74\% & 6\% & 0\% & 20\%\\
    Sync EASGD3 & 0.988 & 1,000 & 7.7s & 10\% & 4\% & 0\% & 79\% & 7\% & 0\% & 14\%\\
  \bottomrule
\end{tabular}
\end{table*}

\subsection{Knights Landing Optimization}
\label{sec:knl_optimization}
Our platform's KNL chip has 68 cores or 72 cores, which is much more than that of a regular CPU chip. To make full use of KNL's computational power, data locality is highly important. Also, we need to make the best use of KNL's cluster mode (Section \ref{sec:knl}) at the algorithm level.
We partition the KNL chip into 4 parts like Quad or SNC-4 mode. The KNL acts like a 4-node NUMA system. In this way, we also replicate the data into 4 parts and each NUMA node gets one part. We make 4 copies of weights and each NUMA node has one copy.
After all the NUMA nodes compute the gradients, we conduct a tree-reduction operation to sum these all gradients. Each NUMA node can get one copy of the gradient sum and use it to update its own weights. In this way, different NUMA nodes do not need to communicate with each other unless they share the gradients. This is a divide-and-conquer method. 
The divide step includes replicating the data and copying the weights. The conquer step is to sum up the gradients from all partitions. This can speedup the algorithm by the faster propagation of gradients.

\begin{figure}[!t]
\centering
\includegraphics[width=3.0in]{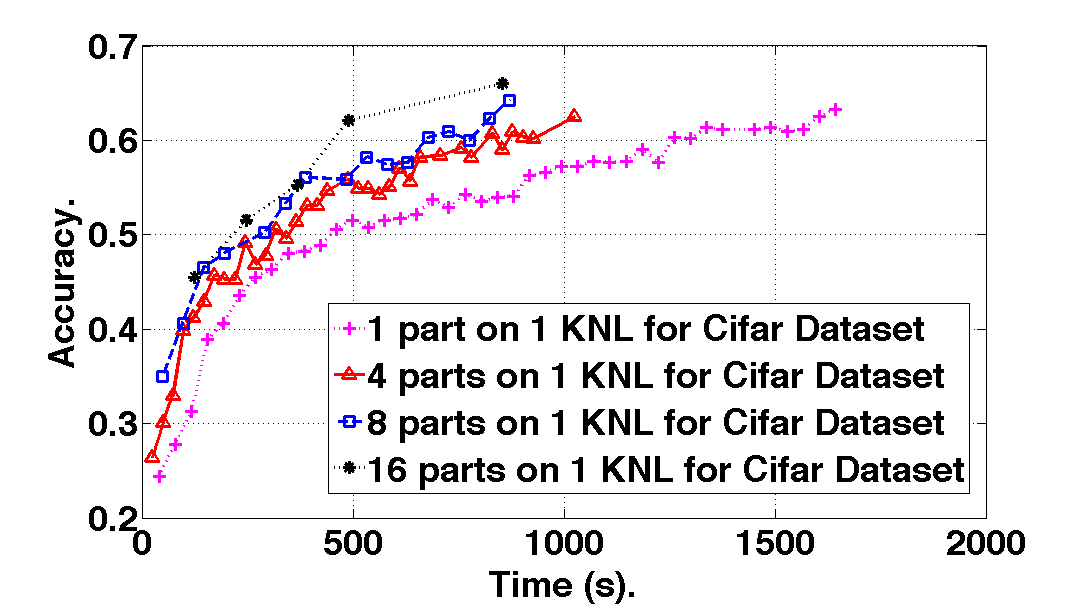}
\caption{Partitioning a KNL chip into group and making each group process one local weight can improve the performance.}
\label{fig:knlpartition}
\end{figure}

In the same way, we can partition the chip into 8 parts, 16 parts, and so on. Let us partition the chip into $P$ parts. The limitation of this method is that the fast memory (cache and MCDRAM) should be able to handle $P$ copies of weight and $P$ copies of data. Figure \ref{fig:knlpartition} shows that this method works for $P \leq 16$ when we use AlexNet to process Cifar dataset. The reason is that the AlexNet is 249 MB and one Cifar data copy is 687 MB. Thus, MCDRAM can hold at most 16 copies of weight and data. Concretely, for achieving the same accuracy (0.625), 1-part case needs 1605 sec, 4-part case needs 1025 sec, 8-part case needs 823 sec, and 16-part case only needs 490 sec. We achieve 3.3$\times$ speedup by copying weight and data to make full use of the fast memory and reduce communication. 

\section{Additional Results and Discussions}

\begin{table*}
  \caption{Weak Scaling Time and Efficiency for ImageNet Dataset}
  \label{tab:weak_scaling}
  \begin{tabular}{l*{9}{c}r}
    \toprule
    Models & 68 cores & 136 cores & 272 cores & 544 cores & 1088 cores & 2176 cores & 4352 cores\\
    \midrule
    GoogleNet (300 Iters Time) & 1533s & 1590s & 1608s & 1641s & 1630s & 1662s & 1674s\\
    GoogleNet (Efficiency) & 100\% & 96.4\% & 95.3\% & 93.4\% & 94.0\% & 92.3\% & 91.6\%\\
    VGG (80 Iters Time)          & 1318s & 1440s & 1482s & 1524s & 1634s & 1679s & 1642s\\
    VGG (Efficiency)          & 100\% & 91.5\% & 89.0\% & 86.5\% & 80.7\% & 78.5\% & 80.2\%\\
  \bottomrule
\end{tabular}
\end{table*}

\begin{figure}[ht]
\centering
\renewcommand{\thesubfigure}{\thefigure.\arabic{subfigure}}
\subfigure[]{\includegraphics[width=2.7in]{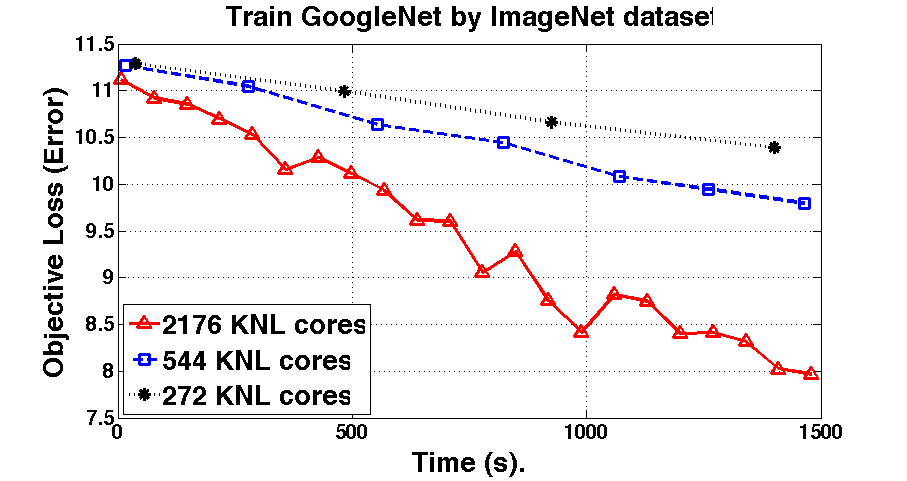}
\label{fig_first_case}}
\subfigure[]{\includegraphics[width=2.7in]{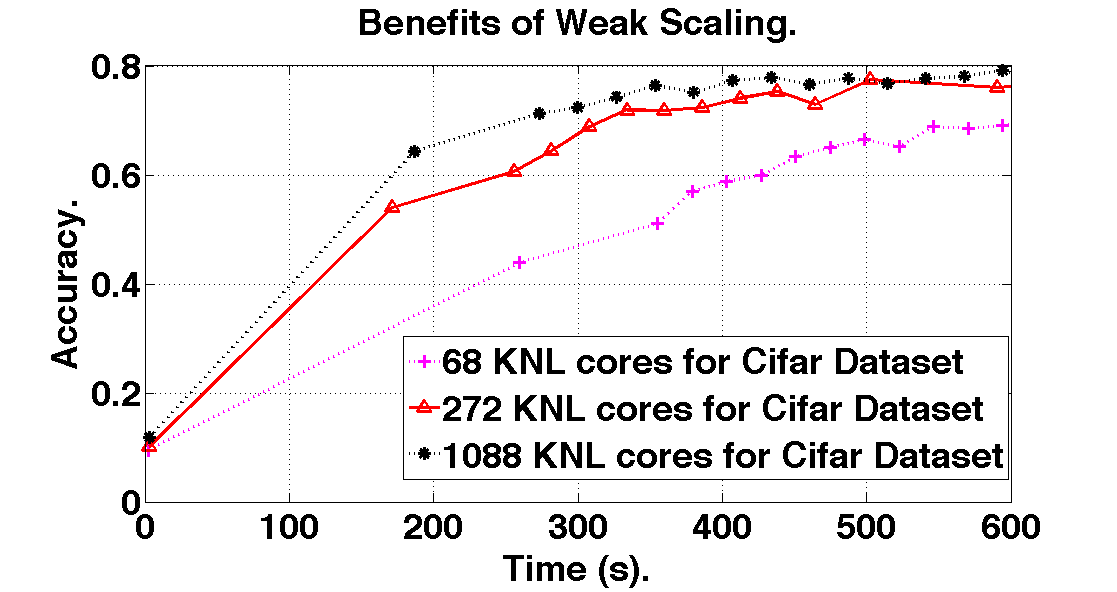}
\label{fig_first_case}}
\caption{The benefits of using more machines and more data: (1) get the target accuracy in a shorter time, and (2) achieve a higher accuracy in a fixed time. Objective Loss means Error (lower is better).}
\label{fig:weak_scaling}
\end{figure}


\subsection{Comparison with Intel Caffe}
Intel Caffe is the state-of-the-art implementation for both single-node and multi-node on Xeon and Xeon Phi platforms. Because this paper is focused on inter-node (distributed) algorithm, we use Intel Caffe for single-node implementation. We only compare with Intel Caffe for scaling because we have the same single-node performance (baseline) with Intel Caffe.

Machine Learning researchers focus on weak scaling because they need higher accuracy when they use more machines and larger datasets in a fixed time (e.g. draw a vertical line in Figure \ref{fig:weak_scaling}).
On the other hand, weak scaling also means getting the target accuracy in a shorter time by using more machines and larger data (e.g. draw a horizontal line in Figure \ref{fig:weak_scaling}).
Figure \ref{fig:weak_scaling} shows the benefit of using more machines and more data. Each node processes one copy of Cifar dataset and the batch size is 64. In the way, we increase the total data size as we increase the number of machines.

For large-scale weak scaling study, we use GoogleNet and VGG to process the ImageNet dataset. Each node has one copy of the ImageNet dataset (240 GB). We increase the number of cores from 68 to 4352. The data size increases as we increases the number of machines. The results of our weak scaling study are shown in Table \ref{tab:weak_scaling}. Compared with Intel's implementation, we have a higher weak scaling efficiency. For GoogleNet on 2176 cores, the weak scaling of Intel Caffe is 87\% while that of our implementation is 92\%.
For VGG on 2176 cores, the weak scaling of Intel Caffe is 62\% while that of our implementation is 78.5\%.

\subsection{The Impact of Batch Size}
When changing the batch size, the users need to change learning rate and momentum at the same time. For small batch sizes (e.g from 32 to 1024), increasing the batch size generally speeds up DNN training because larger batch size makes BLAS functions run more efficiently. Increasing the batch size beyond a threshold (e.g. 4096) generally slows down DNN training because in that regime, the optimization space around minima becomes sharper, requiring more epochs to get the same accuracy \cite{keskar2016large}. For medium batch size (e.g. from 1024 to 4096), the users need to tune batch size, learning rate, and momentum together to speed up the training.

\section{Conclusion}
The current distributed machine learning algorithms are mainly designed for cloud systems. Due to cloud systems' slow network and high fault-tolerance requirement, these methods are mainly asynchronous.
However, asynchronous methods are usually unreproducible, nondeterministic, and unstable.
EASGD has a good convergence property. Nevertheless, the round-robin method makes it inefficient on HPC systems.
In this paper, we designed efficient methods for HPC clusters to speedup deep learning applications' time-consuming training process.
Our methods Async EASGD, Async MEASGD, and Hogwild EASGD are faster than their existing counterpart methods. Sync EASGD or Hogwild EASGD method is the fastest one among our competing methods in this paper.
Sync EASGD3 achieves 5.3$\times$ speedup over original EASGD for the same accuracy (98.8\%) while being deterministic and reproducible. We achieve 91.6\% weak-scaling efficiency, which is higher than the state-of-the-art implementation.

\section{Acknowledgement}
We would like to thank Yang You's 2016 IBM summer intern manager Dr. David Kung and mentor Dr. Rajesh Bordawekar. Yang You finished a multi-node multi-GPU EASGD with less global communication overhead at IBM, which is not included in this paper. We also want to thank Prof. Cho-Jui Hsieh at UC Davis for reading the proof. We used resources of the NERSC supported by the Office of Science of the DOE under Contract No. DEAC02-05CH11231.
Dr. Bulu\c{c} is supported by the Applied Mathematics program of the DOE Office of Advanced Scientific Computing Research under contract number DE-AC02- 05CH11231 and by the Exascale Computing Project (17-SC-20-SC), a collaborative effort of the U.S. DOE Office of Science and the National Nuclear Security Administration. 
Prof. Demmel and Yang You are supported by the U.S. DOE Office of Science, Office of Advanced Scientific Computing Research, Applied Mathematics program under Award Number DE-SC0010200; by the U.S. DOE Office of Science, Office of Advanced Scientific Computing Research under Award Numbers DE-SC0008700; by DARPA Award Number HR0011-12- 2-0016, ASPIRE Lab industrial sponsors and affiliates Intel, Google, HP, Huawei, LGE, Nokia, NVIDIA, Oracle and Samsung. Other industrial sponsors include Mathworks and Cray.
The funding information in \cite{you2015svm} and \cite{you2016asynchronous} maybe also relevant.

\section{Artifact Description Appendix}

\subsection{The Source Code}
We share our source code online\footnote{\href{https://www.cs.berkeley.edu/~youyang/sc17code.zip}{https://www.cs.berkeley.edu/$\sim$youyang/sc17code.zip}}, with everything necessary included.

\subsection{The dataset}
First, due to the limit of file size, we can not upload the datasets. To run our code, the readers need to download the datasets.
For Mnist dataset, the readers can download it from this link\footnote{\label{footnote:code}Mnist dataset is at \href{http://yann.lecun.com/exdb/mnist}{http://yann.lecun.com/exdb/mnist}}.
For Cifar dataset, the readers can download it from this link\footnote{\label{footnote:code}Cifar dataset is at \href{http://www.cs.toronto.edu/~kriz/cifar-10-binary.tar.gz}{http://www.cs.toronto.edu/~kriz/cifar-10-binary.tar.gz}}.
For Imagenet dataset, the readers can download it from this link\footnote{\label{footnote:code}Imagenet dataset is at \href{http://image-net.org/download}{http://image-net.org/download}}.

\subsection{Dependent Libraries}
All our codes are written in C++, and the users need to add -std=c++11 to compile our codes.
For GPU codes, we use CUDA 7.5 and CuBLAS and CuDNN 5.0 libraries. We use Nvidia NCCL for GPU-to-GPU communication. We use MPI for distributed processing on the multi-GPU multi-node system.
For KNL codes, since this paper is focused on inter-node (distributed) algorithm, we use Intel Caffe for single-node implementation. The Intel Caffe depends on Intel MKL for basic linear algebra functions. To install and use Intel Caffe, we install the follow libraries: (1) protobuf/2.6.1, (2) boost/1.55.0, (3) gflags/2.1.2, (4) glog/0.3.4, (5) snappy/1.1.3, (6) leveldb/1.18, (7) lmdb/0.9.18, and (8) opencv/3.1.0-nogui. We use MPI for distributed processing on the KNL cluster.

\subsection{Experimental Systems}
We have two GPU clusters. The first one has 16 nodes. Each node has one Intel E5-1680 v2 3.00GHz CPU and two Nvidia Tesla K80 GPUs. \textcolor{black}{The two halves of the K80 are connected by a PLX Technology, Inc. PEX 8747 48-lane PCIe switch. The nodes are connected by 56 Gbit/s Infiniband.} 
The second cluster has 4 nodes. Each node has one E5-2680 v3 2.50GHz CPU and eight Nvidia Tesla M40 GPUs. \textcolor{black}{Groups of 4 Tesla M100 GPUs are connected with a 96-lane, 6-way PCIe switch. The nodes are connected by 56 Gbit/s Infiniband.}
We test the Knights Landing and CPU algorithm on NERSC's Cori supercomputer, which has 2,004 Haswell CPU nodes (64,128 cores) and 9,688 KNL nodes (658,784 cores in total, 68 cores per node) \textcolor{black}{The CPU version is 16-core Intel Xeon Processor E5-2698 v3 at 2.3 GHz. The KNL version is Intel Xeon Phi Processor 7250 processor with 68 cores per node @ 1.4 GHz. The interconnect for Cori is Cray Aries with Dragonfly topology with 5.625 TB/s global bandwidth (CPU) and 45.0 TB/s global peak bisection bandwidth (KNL).}

\subsection{Running our codes}
After downloading our codes from SC17 submission system and unzipping it. The readers will get two folders: {\bf gpu} and {\bf knl}. 

To run the GPU related codes, the readers need to enter the {\bf gpu} folder. There are eight subfolders in {\bf gpu} folder. Each subfolder corresponds to one method mentioned in this paper. For example, after entering the {\bf mnist\_easgd\_async} subfolder, the readers will find a couple of files. {\bf readubyte.cpp} and {\bf readubyte.h} are for reading the dataset. The algorithm is implemented in {\bf my\_nn.cu} and the readers can use {\bf Makefile} to compile it. After compiling the code, the readers can just execute {\bf run.sh} file to run the program. The {\bf parameter.txt} file defines the neural network structure. The results will be shown on screen and stored in {\bf .out} files.
For running the distributed code, the readers can enter the {\bf mpi\_easgd} subfolder. To compile the code, the readers just need to run {\bf compile.sh} file. Then the readers can execute {\bf run.sh}  file to run the program.

To run the KNL related codes, the readers need to enter the {\bf knl} folder. There are eight subfolders in  {\bf knl} folder. Each subfolder corresponds to one method mentioned in this paper. For example, after entering the {\bf cifar\_average\_sync} subfolder, the readers will find a couple of files. The {\bf solver.prototxt} files define the algorithmic setting (e.g. \# iterations, \# learning rate, and \# testing frequency). The {\bf train\_test.prototxt} files define the structure of neural networks. To compile the code, the readers just need to execute {\bf compile.sh} file. After the compilation, the readers need to submit the program to job management system. We use the slurm workload manager. {\bf myknlrun.sl} file is our submission script. After finishing the job, the readers can use {\bf calacc.cpp} to sum up the accuracy information and {\bf caltime.cpp} to sum up the time information.
\bibliographystyle{ACM-Reference-Format}
\bibliography{sigproc}

\end{document}